\documentstyle[12pt]{article}
\newfont{\ffont}{msym10}                        
\newcommand{\beq}{\begin{equation}}             
\newcommand{\eeq}{\end{equation}}               
\newcommand{\bqry}{\begin{eqnarray}}            
\newcommand{\eqry}{\end{eqnarray}}              
\newcommand{\bqryn}{\begin{eqnarray*}}          
\newcommand{\eqryn}{\end{eqnarray*}}            
\newcommand{\preprint}[1]{\begin{table}[t]      
            \begin{flushright}                  
            \begin{large}{#1}\end{large}        
            \end{flushright}                    
            \end{table}}                        
\newcommand{\PD}[2]                             
    {\frac{\partial^{#2}}{\partial #1^{#2}}}    
\begin{document}
\preprint{TAUP-2048-93 \\  }
\title{Equilibrium Relativistic Mass Distribution}
\author{\\ L. Burakovsky\thanks {Bitnet:BURAKOV@TAUNIVM.} \
and L.P. Horwitz\thanks
  {Bitnet:HORWITZ@TAUNIVM. Also at Department of Physics, Bar-Ilan
University, Ramat-Gan, Israel  } \\ \ }
\date{School of Physics and Astronomy \\ Raymond and Beverly Sackler
Faculty of Exact Sciences \\ Tel-Aviv University,
Tel-Aviv 69978, ISRAEL}
\maketitle
\begin{abstract}
The relativistic Maxwell-Boltzmann distribution for the system of $N$
events with motion in space-time parametrized by an invariant
``historical time'' $\tau $ is considered without the simplifying
approximation $m^2\cong M^2$,
where $M$ is a given intrinsic property of
the events. The relativistic mass distribution is obtained and the
average values of $m$ and $m^2$ are calculated. The average value of
the energy in nonrelativistic limit gives a correction of the order of
10\% to the Dulong-Petit law. Expressions for the pressure and the
density of events are obtained and the ideal gas law is recovered.
\end{abstract}

\bigskip
\section{Introduction}

This article is the natural continuation of the work of Horwitz, Shashoua
and Schieve \cite{HorShaSch}, which in the present text will be cited
as $I$. In that work a derivation of a manifestly covariant
relativistic Boltzmann equation for the system of $N$ events was made in
the framework of a manifestly covariant classical and quantum
mechanics \cite{rqm}. These events,
considered as the fundamental dynamical objects of the theory, move in an
$8N$-dimensional phase space. Their motion is parametrized by a
continuous Poincar\'{e} invariant parameter $\tau $ called the
``historical time''.
The collection of events (called ``concatenation'' \cite {conc}) along
each world line corresponds to a $particle$ in the usual sense; e.g.,
the Maxwell conserved current is an integral over the history of the
charged event \cite{Jack} . Hence the evolution of the state of the
$N$-event system describes the history in space and time of an
$N$-particle system.  The relativistic Boltzmann
equation was used to prove the $H$-theorem for evolution in $\tau$. In
the equilibrium limit a covariant form of the Maxwell-Boltzmann
distribution was obtained. Since this distribution
is the distribution of the 4-momenta of the events, $m^2=-p^2=
-p^{\mu }p_{\mu }$ is a random variable in a relativistic ensemble.
In order to obtain a simple analytic result
the authors restricted themselves to a narrow mass shell
$p^{2}=-m^{2}\cong -M^{2}$ where $M$ is a given fixed parameter, with
the dimension of mass (an intrinsic property of the events),
assumed to be the same for all the events of the system.
The results obtained in this approximation are in agreement
with the well-known results of Synge \cite{Syn}
from an on-mass-shell relativistic kinetic theory.

As Hakim \cite{Hak} has remarked, some problems of relativistic
statistical theory require consideration of
relativistic particles endowed with variable masses. Thus, in
$statistical\;cosmology$ the universe is constituted of a gas of
particles with equal masses. The particles are, in turn, considered as
being galaxies and possibly clusters of galaxies or stars. However,
galaxies and clusters of galaxies\footnote{A number of clusters, such as
Coma or Virgo,
contains large numbers of galaxies of unequal masses.} do not appear to
have identical masses. Therefore, a consistent approach to statistical
cosmology as well as the statistical treatment of clusters of galaxies
themselves should involve a $mass\;distribution.$

Moreover, in the $statistical\;bootstrap\;model$ of Hagedorn \cite{Hag}
and Frautschi \cite{Fra} for multiple production of particles in high
energy reactions a mass spectrum of the asymptotic form $\rho (m)
\sim cm^a\exp(bm)$ (where $a,b,c$ are constants) lies in the
foundations
of their theory and gives a good qualitative agreement with relevant
experiments in high-energy physics. Miller and Suhonen \cite{M&S}
have discussed a possible correlation of the grand canonical
distribution function of ref.\cite{HorSchPi}, characterized
by the mass fluctuations, with the hadronic spectrum of
the Hagedorn-Frautschi form.

In this paper, we study such a relativistic system without the
simplifying approximation $m^{2}\cong M^{2}$, i.e., we consider a
relativistically covariant Maxwell-Boltzmann distribution with mass
parameter $m$ taking any value within the range $0\leq m<\infty $. We
obtain the relativistic mass distribution by integration of the Maxwell-
Boltzmann distribution over angular and hyperbolic angular variables.
Calculation of average value of
the energy and study of its nonrelativistic approximation gives a
relativistic correction of the order of $10\%$ to the Dulong-Petit
law for the free-particle gas assuming that the mass distribution remains
valid in this approximation of our theory. We remark that,
as is well known \cite{LevyL}, the structure of the Galilean group,
the symmetry of non-relativistic systems,
implies that the mass of a particle must be a constant intrinsic
property. We recognize, however, that the applicability of the Galilean
group is an idealization of a world which seems to be more correctly
described by the Poincar\'{e} group. The result that we have found
follows from equilibrium thermodynamics without imposing the geometrical
restriction of the precise Galilean group to an infinitely sharp mass
shell.

By examining the energy-momentum tensor we obtain expressions for the
pressure and the density of events and recover the ideal gas law
previously obtained in $I$.

Since there was no fundamental theory available to Hakim \cite{Hak2},
his development of the theory followed the
phenomenological methods initiated by Synge [5].
In this way he obtained a relativistic statistical mechanics containing
many useful results \cite{HaM}. In particular, he used the
J\"{u}ttner-Synge form (coinciding with the sharp-mass or low-temperature
limit of the theory which we discuss here) to obtain a mass distribution
[6]. As we shall see, this construction leads to a
normalization of the distribution function, which differs from ours.
The theoretical framework
which we use, permitting a derivation of the relativistic Boltzmann
equation from first principles, provides a condition on the equilibrium
ensemble which leads to a well-defined mass distribution (independent of
the J\"{u}ttner-Synge result, although consistent with it), and a
normalization condition consistent with a quantum-mechanical positive
definite density.

\section{Relativistic mass distribution}

We begin with the Maxwell-Boltzmann distribution function which we write
down in the form (differing from one obtained in   [1,(3.14)]
by the signs in the exponent\footnote{These signs do not influence the
requirements for the Maxwell-Boltzmann distribution because
the logarithm of the distribution function
$$\ln f_{0}(q,p)=A(p+p_{c})^{2}+\ln C(q)$$
possesses the property [1,(3.10)]
$$\ln f_{0}(q,p)=\chi _{1}(q,p)+\chi _{2}(q,p)+...,$$
where the quantities $$\chi _{i}(q,p_{1})+\chi _{i}(q,p_{2})$$ are
conserved in collisions. For the sharp mass approximation $m^2\cong M^2$
the difference between the two expressions manifests itself only in the
normalization factor.}; we also use the metric $g^{\mu \nu }=(-,+,+,+)$
and $q\equiv q^\mu ,\;p\equiv p^\mu )$
\beq
f_{0}(q,p)=C(q)e^{A(q)(p+p_{c})^{2}},\!\; A(q)>0.
\eeq

The function (1) must be normalized, according to [1,(2.14)], as
\beq
n(q)=\int f_{0}(q,p)d^{4}p=C(q)\int d^{4}p\exp \{ A(p+p_{c})^{2}\} ,
\eeq
where $n(q)$ is the total number of events per unit space-time volume
in the system in the neighborhood of the point $q$. By introducing
hyperbolic variables [1,(3.16),(3.17)]
$$\Omega ^4: m\geq 0,0\leq \theta \leq \pi ,
0\leq \varphi <2\pi ,-\infty <\beta <\infty ,$$
we can rewrite (2) as follows:
\beq
n(q)=C(q)e^{-Am^{2}_{c}}
\int _{\Omega ^{4}}m^{3}\sinh^{2}\beta \sin\theta
dmd\beta d\theta d\varphi e^{-Am^{2}-2Am_{c}m\cosh \beta }.
\eeq
After some
calculations \cite{GrRy} we obtain the normalization relation
\beq
n(q)=C(q)e^{-Am^2_c}\frac {\pi }{2A^2}\Psi (2,2;Am^2_c).
\eeq
Here $\Psi (a,b;z)$ is the confluent hypergeometrical function of $z$
(ref.[15], p.257, sec.6.6).
After integration over angular and hyperbolic
angular variables we obtain from (3) the expression
\beq
n(q)=C(q)\frac {2\pi e^{-Am_c^2}}{Am_c}
\int _{0}^{\infty }dmm^2e^{-Am^2}K_1 (2Am_cm),
\eeq
from which we identify the mass distribution function
\beq
f(m)=Dm^2e^{-Am^2}K_1(2Am_c m),
\eeq
where
\beq
D^{-1}=\frac {m_c}{4A}\Psi (2,2;Am_c^{2}),
\eeq
so that $f(m)$ is normalized according to
\beq
\int ^{\infty }_0 f(m)dm=1.
\eeq
In (5) and (6) $K_1$ is the Bessel function of the third kind
(imaginary argument), where, in general,
$$K_{\nu }(z)=\frac{\pi i}{2}e^{\pi i\nu /2}H_{\nu }^{(1)}(iz).$$

With the help of this distribution one can obtain the local average
value of an arbitrary function of mass $\phi (m)$:
\beq
\langle \phi (m)\rangle _q=\int _0^\infty \phi (m)f(m)dm.
\eeq
Let us obtain, for example, the local average values of mass and
mass squared in relativistic gas which represent the first two moments
of the distribution (6):
\beq
\langle m\rangle _q=\frac {3\pi }{8A^{\frac{1}{2}}}
\frac {\Psi (\frac{5}{2},2;Am_c^2)}{\Psi (2,2;Am_c^2)},
\eeq
\beq
\langle m^2\rangle _q=\frac {2}{A}
\frac {\Psi (3,2;Am_c^2)}{\Psi (2,2;Am_c^2)}.
\eeq
More generally,
\beq
\langle m^{\ell }\rangle _q=
\Gamma (\frac{\ell }{2}+1)\Gamma (\frac {\ell }{2}+2)
A^{-\frac{\ell }{2}}\frac {\Psi (\frac{\ell }{2}+2,2;Am_c^2)}
{\Psi (2,2;Am_c^2)}.
\eeq
Now, as in $I$, we define absolute temperature through the
relation [1,(3.26)]
\beq
2Am_c=\frac {1}{k_BT},
\eeq
which implies that in thermal equilibrium $Am_c$ is independent of
$q$. Hence
\beq
\langle m\rangle =\frac {3\pi }{8}\sqrt {2m_ck_BT}
\frac {\Psi (\frac{5}{2},2;\frac{m_c}{2k_BT})}
{\Psi (2,2;\frac{m_c}{2k_BT})}.
\eeq
In the limit $T\rightarrow 0$ it follows from the asymptotic
formula for $z\rightarrow \infty $ \cite{conf2}
\beq
\Psi (a,b;z)\sim z^{-a}[1+\sum _{n=1}^\infty (-1)^n
\frac {a(a+1)\cdots (a+n-1)(1+a-b)(2+a-b)\cdots (n+a-b)}{n!z^n}]
\eeq
that
\beq
\langle m\rangle \cong \frac {3\pi }{4}k_BT.
\eeq
One can also obtain in this limit
\beq
\langle m^2\rangle \cong 8(k_BT)^2.
\eeq

Let us now calculate the first two moments of the distribution (1):
\beq
\langle p^{\mu }\rangle _q=
\frac {\int d^4pp^{\mu }e^{A(p+p_c)^2}}{\int d^4pe^{A(p+p_c)^2}},
\eeq
\beq
\langle p^{\mu }p^{\nu }\rangle _q=
\frac {\int d^4pp^{\mu }p^{\nu }e^{A(p+p_c)^2}}
{\int d^4pe^{A(p+p_c)^2}}.
\eeq
If we introduce the ``free energy'' of the relativistic gas through
the relation
\beq
\int d^4pe^{A(p+p_c)^2}=e^{-AF},
\eeq
we will obtain the following expressions:
\beq
\langle p^{\mu }\rangle _q=p^{\mu }_c(F^{\prime }-1),
\eeq
\beq
\langle p^{\mu }p^{\nu }\rangle _q=
p^{\mu }_cp^{\nu }_c\left[(F^{\prime }-1)^2-\frac{F^{\prime \prime }}
{A}\right]+g^{\mu \nu }\frac{F^{\prime }-1}{2A},
\eeq
\beq
F^{\prime }\equiv \frac {\partial F}{\partial m_c^2}.
\eeq
The value of $F$, obtained from (20) and (4), is
$$F=-\frac {1}{A}\ln [\frac{\pi }{2A^2}e^{-Am_c^2}\Psi (2,2;Am_c^2)].$$
A simple calculation with the help of the relation (ref.[15], p.257,
sec.6.6) $$\frac{d}{dz}\Psi (a,b;z)=-a\Psi (a+1,b+1;z)$$ gives
\beq
\langle p^{\mu }\rangle _q=2p_c^{\mu }
\frac {\Psi (3,3;Am_c^2)}{\Psi (2,2;Am_c^2)},
\eeq
\beq
\langle p^{\mu }p^{\nu }\rangle _q=
6\frac {\Psi (4,4;Am_c^2)}{\Psi (2,2;Am_c^2)}p_c^{\mu }p_c^{\nu }+
\frac {g^{\mu \nu }}{A}
\frac {\Psi (3,3;Am_c^2)}{\Psi (2,2;Am_c^2)}.
\eeq
As in $I$, we make a Lorentz transformation to the local average
motion rest frame moving with the relative velocity
$${\bf u}=\frac{{\bf p}_c}{E_c},$$ in order to obtain the local
energy density.

The rest frame energy is $$\langle E^{\prime }\rangle _q=\frac
{\langle E\rangle _q-{\bf u}\cdot {\bf p}}{\sqrt {1-{\bf u}^2}},$$
so that
\beq
\langle E^{\prime }\rangle _q=2m_c
\frac {\Psi (3,3;Am_c^2)}{\Psi (2,2;Am_c^2)}.
\eeq
Using the asymptotic formula (15) and expression (16), we have
for $T\rightarrow 0$,
\beq
\langle E^{\prime }\rangle -\langle m\rangle \cong (4-
\frac {3\pi }{4})k_BT=\gamma \frac {3}{2}k_BT,
\eeq
\beq
\gamma =\frac {16-3\pi }{6}\approx 1.1.
\eeq
This result differs from the nonrelativistic result $\frac {3}{2}
k_BT$. The coefficient $\gamma $ represents a relativistic correction,
determined by the relativistic mass distribution, to the classical
value $3\over 2$.
The existence of such corrections was first shown by Horwitz, Schieve
and Piron [10] in their work on study of relativistic Gibbs ensembles
even in the case of small fluctuations of mass over its sharp value
$M_0$.

In the limit $T\rightarrow \infty $ we use the asymptotic formulas
for $z\rightarrow 0 $ (ref.[15], p.262, sec.6.8)
\beq
\Psi (a,b;z)=\frac {\Gamma (b-1)}{\Gamma (a)}z^{1-b}+O(\mid z\mid^
{Re \!\; b-2}),\; Re \!\; b\geq 2,\; b\neq 2,
\eeq
\beq
\Psi (a,b;z)=\frac {\Gamma (b-1)}{\Gamma (a)}z^{1-b}+O(\mid \ln z\mid )
,\; b=2,
\eeq
and obtain
\beq
\langle E^{\prime }\rangle \cong 2k_B T,
\eeq
the result obtained also in $I$.

In this limit one can find from (10),(11)
\beq
\langle m\rangle  \cong \sqrt{\frac{\pi m_c k_B T}{2}},
\eeq
\beq
\langle m^2\rangle \cong 2m_c k_BT.
\eeq
To obtain the pressure and density distributions in our ensemble,
as in $I$, we study the $particle$ energy-momentum
tensor defined by the $R^4$ density
\beq
T^{\mu \nu }(q)=\sum _{i}\int d\tau \frac {p^{\mu }_i p^{\nu }_i }{M}
\delta ^4 (q-q_i(\tau )).
\eeq
Using the result of the previous work [1,(3.37)]
\beq
\langle T^{\mu \nu }(q)\rangle _q =T_{\triangle V}\int d^4 pf_0 (q,p)
\frac {p^\mu p^\nu }{M}
\eeq
and the expression (25) for $\langle p^\mu p^\nu \rangle _q $, we obtain
\beq
\langle T^{\mu \nu }(q)\rangle _q =\frac {T_{\triangle V}n(q)}{M}\left[
\frac{g^{\mu \nu }}
{A}\frac {\Psi (3,3;Am^2_c )}{\Psi (2,2;Am^2_c)}+6p^\mu _c
p^\nu _c \frac {\Psi (4,4;Am^2_c)}{\Psi (2,2;Am^2_c )}\right].
\eeq
In this expression $T_{\triangle V}$ is the average passage interval in
$\tau $ for the events which pass through the small four-volume
$\triangle V $ over the point $q$ of $R^4$.

The formula for the stress-energy tensor of a perfect fluid has the form
[1,(3.39)]
\beq
\langle T^{\mu \nu }(q)\rangle _q =pg^{\mu \nu }-(p+\rho )\frac {\langle
p^{\mu }\rangle _q \langle p^{\nu }\rangle _q }
{\langle p^{\lambda }\rangle _q \langle p_{\lambda }\rangle _q},
\eeq
where $p$ is the pressure and $\rho $ is the density of energy at $q$.

According to  (24),
\beq
\frac{\langle p^{\mu }\rangle _q }
{\sqrt{-\langle p^{\lambda }\rangle _q \langle p_{\lambda }\rangle _q }}
=\frac{p^{\mu }_c}{m_c},
\eeq
hence
\beq
p=\frac{T_{\triangle V}n(q)}{AM}\frac{\Psi (3,3;Am^2_c)}
{\Psi (2,2;Am^2_c)}
\eeq
and
\beq
p+\rho =\frac{6T_{\triangle V}n(q)m^2_c}{M}\frac{\Psi (4,4;Am^2_c)}
{\Psi (2,2;Am^2_c)}.
\eeq
To interpret these results, as in $I$, we should
calculate the average (conserved) $particle$
four-current, which has the microscopic form
\beq
J^{\mu }(q)=\sum _i \int \frac {p^{\mu }_i}{M}\delta ^4 (q-q_i(\tau ))d
\tau .
\eeq
Using the result of $I$ [(3.45),(3.59)]
\beq
\langle J^{\mu }(q)\rangle _q =T_{\triangle V}\int d^4 p\frac{p^{\mu }}
{M}f_0 (q,p)
\eeq
and expression (24) for $\langle p^{\mu }\rangle _q$, we obtain
\beq
\langle J^{\mu }(q)\rangle _q =
\frac{2T_{\triangle V}n(q)}{M}p^{\mu }_c \frac{\Psi (3,3;Am^2_c)}
{\Psi (2,2;Am^2_c)}.
\eeq
In the local rest frame $p^{\mu }_c =(m_c,{\bf 0})$,
\beq
\langle J^0(q)\rangle _q=\frac{2T_{\triangle V}n(q)m_c}
{M}\frac{\Psi (3,3;Am^2_c)}{\Psi (2,2;Am^2_c)}.
\eeq
Defining the density of $paricles$ per unit space volume
[1,(3.48)] as
\beq
N_{0}(q)=\langle J^{0}(q)\rangle _q,
\eeq
we obtain the ideal gas law [1,(3.49)]
\beq
p=\frac{N_0}{2Am_c}=N_0k_B T.
\eeq

Now we shall show how the general expression for the distribution
function (6) can be simplified in two limiting cases $T\rightarrow 0$
and $T\rightarrow \infty .$

1. $T\rightarrow 0$. In this case we can see from (17) and (27) that
$\langle m^2\rangle \sim (k_BT)^2\ll \langle E^{\prime}\rangle m_c
\sim m_c(k_BT).$ We therefore
neglect $m^2=-p^2$ in comparison to $2App_c$ in
the exponent of the Maxwell-Boltzmann distribution (1); hence,
we begin with
\beq
f_0^{(1)}(p,q)\cong C(q)e^{-Am_c^2}e^{2App_c},
\eeq
and after integration and normalization obtain
\beq
f^{0}(m)=\frac{(2Am_c)^3}{2}m^2K_1(2Am_cm).
\eeq
It is easy to see that the distribution function (48) produces
results for $\langle m\rangle $ and $\langle m^2\rangle $ in this
limit in agreement with (16) and (17):
$$\langle m\rangle ^{0}=\frac{3\pi }{4}k_BT,\;
\langle m^2\rangle ^{0}=8(k_BT)^2.$$ More generally,
\beq
\langle m^{\ell }\rangle ^{0}=\Gamma (\frac{\ell }{2}+1)
\Gamma (\frac{\ell }{2}+2)(Am_c)^{-\ell }=\Gamma (\frac{\ell }{2}+1)
\Gamma (\frac{\ell }{2}+2)(2k_BT)^{\ell },
\eeq
which coincides with the low-temperature limit of (12).
Similarly, if we introduce the ``free energy'' for the distribution (47),
we obtain
\beq
\langle p^{\mu }\rangle ^{0}=\frac{p_c^{\mu }}{m_c}4k_BT,
\eeq
which coinsides with the low-temperature limit of (24).

2. $T\rightarrow \infty .$ It is seen from (31) and (33) that
$\langle m^2\rangle $ and $\langle E^{\prime}\rangle m_c$ are of the
same order in $T.$ But the argument of the function
$K_1$ in the expression of the mass distribution function (6) is of the
lower order of $T$ in this limit
[(32)], i.e., $2Am_cm\sim m_c^{\frac{3}{2}}(k_BT)^{\frac{1}
{2}}.$ Since we use the asymptotic formula \cite{AS}
$$K_{\nu}(z)\sim \frac{1}{2}\Gamma (\nu )\left(\frac{z}{2}\right)
^{-\nu },\;z\rightarrow 0,$$ which gives $K_1(2Am_cm)\cong 1/2Am_cm,$ and
obtain the (normalized) distribution
\beq
f^{\infty }(m)=2Ame^{-Am^2}.
\eeq
One can easy check that the distribution (51) gives results
for $\langle m\rangle $ and $\langle m^2\rangle $ in this limit
in agreement with (32) and (33): $$\langle m\rangle ^{\infty }=
\sqrt {\frac{\pi m_ck_BT}{2}},\;\langle m^2\rangle ^{\infty }
=2m_ck_BT.$$ More generally,
\beq
\langle m^{\ell }\rangle ^{\infty }=\Gamma (\frac{\ell }{2}+1)
A^{-\frac{\ell }{2}}=
\Gamma (\frac{\ell }{2}+1)(2m_ck_BT)^{\frac{\ell }{2}},
\eeq
which coincides with the high-temperature limit of (12).

In this way one can also obtain
\beq
\langle p^{\mu }\rangle ^{\infty }=\frac{p_c^{\mu }}{m_c}
2k_BT,
\eeq
the high-temperature limit of (24).

Summarizing, we can write down the expression for the mass distribution
function, taking into account the limiting cases considered:
\beq
f(m)=\left\{ \begin{array}{ll}
\{(2Am_c)^3/2\}m^2K_1(2Am_cm), & T\rightarrow 0 \\     &    \\
\{4A/m_c\Psi (2,2;Am_c^2)\}e^{-Am^2}m^2K_1(2Am_cm), &
{\rm intermediate\;\;case} \\    &    \\
2Ame^{-Am^2}, & T\rightarrow \infty
\end{array} \right.
\eeq

Synge [5], following phenomenological methods in his
study of an on-mass-shell
relativistic ensemble, used the normalization relation (the only
normalization relation available to him)
for the equilibrium distribution function for such a system\footnote{in
his notation, $2Ap_c^{\mu }=\xi ^{\mu }.$},
\beq
f_0(q,p)=G(q)e^{2App_c},
\eeq
called the ``J\"{u}ttner-Synge distribution function'',
\beq
\int f_0(q,p)d^4p=G(q)\int d^4pe^{2App_c}=N_0(q),
\eeq
where $N_0(q)=\langle J_0(q)\rangle _q$ is the density of $particles$
per unit space-volume (45), the $0$-component of the average
(conserved) $particle$ four-current; in contrast to the normalization
relation (56), we have the condition (2),
$$\int f_0(q,p)d^4p=n(q),$$ where $n(q)$ is the density of $events$ per
unit space-time volume, a quantum-mechanical positive-definite density,
which follows from the theory discussed in $I.$

In the theory of Synge $$\langle J^{\mu }(q)\rangle _q=\int f_0(q,p)
\frac{p^{\mu }}{m}d^4p=\frac{G(q)}{m}\frac{\partial \Phi }{\partial
(2Ap_c^{\mu })},$$ where $m$ is a (constant) mass of the particles of the
system and (ref.[5], p.35)
\beq
\Phi =\int e^{2App_c}d^4p=\frac{4\pi mK_1(2Am_cm)}{2Am_c}.
\eeq
In this way he obtained the normalization relation (ref.[5], p.36)
\beq
N_0=\frac{4\pi Gm^2K_2(2Am_cm)}{2Am_c},
\eeq
differing from the one obtained in $I$ [1,(3.18)]
in the mass-sharp-approximation $m^2
\cong M^2$ for the same distribution function (55) within the framework
of the theory discussed in $I$:
\beq
n_0(q)=\frac{4\pi GM^2K_1(2AM_cM)2\triangle m}{2AM_c},
\eeq
where $G=C(q)\exp\{A(M+M_c)^2\}$ and
$\triangle m$ is the fluctuation of mass around its sharp value $M.$

Hakim introduced variation of mass into the J\"{u}ttner-Synge
distribution and, using the normalization condition (58),
found the mass distribution [6,(3.3)]
\beq
f(m)=\frac{2(2Am_c)^3}{3\pi }m^2K_2(2Am_cm).
\eeq
This distribution differs from our low-temperature limit (48)
$$f^0(m)=\frac{(2Am_c)^3}{2}m^2K_1(2Am_cm),$$ because it is obtained by
assuming a different normalization of the initial
J\"{u}ttner-Synge distribution. This is why the results of Hakim differ
quantitatively from the results obtained within the framework of the
theory discussed in the present paper in the low-temperature limit.
\section{Concluding remarks}

We have considered an equilibrium relativistic ensemble, described by
the equilibrium relativistic Maxwell-Boltzmann distribution, with
variable mass. We have found that the equilibrium state of such a system
is characterized by a well-defined mass distribution (consistent with
the J\"{u}ttner-Synge one in the sharp-mass or low-temperature limit),
following directly from the Maxwell-Boltzmann distribution, by
integration
over angular and hyperbolic angular variables and satisfying a
normalization condition consistent with a quantum-mechanical positive
definite density.

The results of Hakim, who introduced variation of mass into the
equilibrium relativistic J\"{u}ttner-Synge distribution and found a
relativistic mass distribution [6], differ quantitatively from the
results obtained within the framework of the theory discussed in the
present paper in the low-temperature limit (e.g., he did not obtain a
relativistic correction to the Dulong-Petit law) because the initial
J\"{u}ttner-Synge distribution, from which the mass distribution of Hakim
is obtained, has a normalization differing from ours, based on the
counting of world lines ($particles$) rather than $events.$

The relativistic mass distribution obtained in this paper may find
applications in relativistic cosmology and astrophysics (when dealing
with clusters of galaxies or stars), in the statistical mechanics of
hadrons, in physics of relativistic plasmas and relativistic liquids.

In the present paper as well as in the previous one ($I$)
the simplest case of relativistic equilibrium is treated.
The nonequilibrium case may be treated separately, e.g., by application
of the Grad method of moments adopted for the relativistic theory.

The case of indistinguishable particles will be also treated separately
\cite{BH}, in order to obtain their equilibrium relativistic distribution
function and the
corresponding equilibrium relativistic mass distribution.

\end{document}